\documentclass[sn-mathphys,Numbered]{sn-jnl}

\usepackage{graphicx}%
\usepackage{multirow}%
\usepackage{amsmath,amssymb,amsfonts}%
\usepackage{amsthm}%
\usepackage{mathrsfs}%
\usepackage[title]{appendix}%
\usepackage{xcolor}%
\usepackage{textcomp}%
\usepackage{manyfoot}%
\usepackage{booktabs}%
\usepackage{algorithm}%
\usepackage{algorithmicx}%
\usepackage{algpseudocode}%
\usepackage{listings}%

\begin{document}

\title[Article Title]{Modeling of cosmic rays and near-IR photons in aluminum KIDs}


\author*[1,2]{\fnm{Elijah} \sur{Kane}}\email{ekane@caltech.edu}

\author[1,2]{\fnm{Chris} \sur{Albert}}

\author[1,2]{\fnm{Ritoban} \sur{Basu Thakur}}

\author[1,2]{\fnm{Charles (Matt)} \sur{Bradford}}

\author[3]{\fnm{Nicholas} \sur{Cothard}}

\author[2]{\fnm{Peter} \sur{Day}}

\author[1,2]{\fnm{Logan} \sur{Foote}}

\author[1]{\fnm{Steven} \sur{Hailey-Dunsheath}}

\author[1,2]{\fnm{Reinier} \sur{Janssen}}

\author[2]{\fnm{Henry (Rick)} \sur{LeDuc}}

\author[1]{\fnm{Lun-Jun (Simon)} \sur{Liu}}

\author[1,2]{\fnm{Hien} \sur{Nguyen}}

\author[1,2]{\fnm{Jonas} \sur{Zmuidzinas}}

\affil*[1]{\orgname{California Institute of Technology}, \orgaddress{\street{1200 E California Blvd}, \city{Pasadena}, \postcode{91125}, \state{California}, \country{USA}}}

\affil[2]{\orgname{Jet Propulsion Laboratory, California Institute of Technology}, \orgaddress{\street{4800 Oak Grove Dr}, \city{Pasadena}, \postcode{91109}, \state{California}, \country{USA}}}

\affil[3]{\orgname{NASA Goddard Space Flight Center}, \orgaddress{\street{8800 Greenbelt Road}, \city{Greenbelt}, \postcode{20771}, \state{Maryland}, \country{USA}}}


\abstract{The PRobe far-Infrared Mission for Astrophysics (PRIMA) is working to develop kinetic inductance detectors (KIDs) that can meet the sensitivity targets of a far-infrared spectrometer on a cryogenically cooled space telescope. An important ingredient for achieving high sensitivity is increasing the fractional-frequency responsivity. Here we present a study of the responsivity of aluminum KIDs fabricated at the Jet Propulsion Laboratory. Specifically, we model the KID’s temporal response to pair-breaking excitations in the framework of the Mattis-Bardeen theory, incorporating quasiparticle recombination dynamics and the pair-breaking efficiency. Using a near-IR laser, we measure time-resolved photon pulses and fit them to our model, extracting the time-resolved quasiparticle density and the quasiparticle recombination lifetime. Comparing the fit to the known energy of the laser provides a measurement of the pair-breaking efficiency. In addition to photon-sourced excitations, it is important to understand the KID’s response to phonon-sourced excitations from cosmic rays. We measure the rate of secondary cosmic rays detected by our devices, and predict the dead time due to cosmic rays for an array in L2 orbit. This work provides confidence in KIDs’ robustness to cosmic ray events in the space environment.}




\maketitle

\section{Introduction}\label{sec1}

The PRobe far-Infrared Mission for Astrophysics (PRIMA) is a concept for a space telescope mission in the far-infrared which aims to study the cosmic history of nucleosynthesis, star formation, and supermassive black hole growth. To achieve this, PRIMA will perform imaging and spectroscopic observations across the $\lambda = 24-235\mu m$ range, using actively cooled optics. Our particular focus is a dispersive direct-detection spectrograph with resolving power between 100 and 200.

The sensitivity goal for PRIMA's FIRESS spectrometer is a detector NEP of $\sim 1\times 10^{-19}$ W Hz$^{-1/2}$, in order to match the fundamental sensitivity limit set by photon shot noise from zodiacal dust. PRIMA will use kinetic inductance detectors (KIDs) to meet this goal. In this paper, we study the pair-breaking efficiency $(\eta_{pb})$ and the quasiparticle lifetime $(\tau_{qp})$, two material parameters of the KID which affect the sensitivity. Understanding the values of these parameters is motivated by the fact the NEP from two-level systems scales as $(\eta_{pb}^{-1}\tau_{qp}^{-1})$ and the NEP from generation and recombination of quasiparticles scales as $(\eta_{pb}^{-1} \tau_{qp}^{-1/2})$ at low temperatures \cite{JZ12}. In Sec. \ref{sec:background}, we develop a response model of a KID to pair-breaking excitations in the Mattis-Bardeen framework. In Sec. \ref{sec:photon}, we apply this model to measure the pair-breaking efficiency and quasiparticle lifetime in an aluminum KID on a small array designed for $\lambda=25\mu m$ observations, using excitations sourced by near-IR photons. In Sec. \ref{sec:deadtime}, we measure the cosmic ray rates of KIDs on a large array designed for $\lambda=210\mu m$ observations. For NEP measurements of the short- and long-wavelength arrays respectively, we refer the reader to our companion manuscripts \cite{nick}, \cite{logan}.

\section{KID response model}
\label{sec:background}

We use equations derived from Mattis-Bardeen theory to relate the number of quasiparticles in the KID inductor to the quality factor $Q_{MB}$ and fractional detuning $\delta x_{MB}$ of the resonance \cite{shd2018}:
\begin{equation}
    \label{eq:MB}
    \delta x_{MB} = -\frac{\alpha_{kin}\gamma S_2}{4N_0 \Delta_0} \delta n_{qp}, \quad \quad Q_{MB}^{-1} = \frac{\alpha_{kin}\gamma S_1}{2N_0 \Delta_0} n_{qp}.
\end{equation}
Above, $\delta n_{qp}$ is a perturbation to the equilibrium quasiparticle density $n_{qp}$. $\alpha_{kin}$ is the kinetic inductance fraction, $\gamma=1$ for thin films, $N_0$ is the single-spin density of states at the Fermi level, and $\Delta_0$ is the gap energy. We adopt standard expressions for $S_1$ and $S_2$ (\cite{JZ12}, Eqs. 71 and 72).

In our homodyne readout scheme, a probe tone is transmitted through the feedline which couples to the KID, and the output signal is mixed with the input to determine the voltage transmission $S_{21}$. The transmission, accounting for possible impedance mismatch between the input and output transmission lines \cite{Zmis}, can be expressed as follows:
\begin{equation}
    \label{eq:S21}
    S_{21} = 1-\frac{1}{1+2jy}\frac{Q_r}{Q_c\cos\phi}e^{j\phi}
\end{equation}
$Q_r$ is the loaded quality factor, which is composed of three terms: $Q_r^{-1} = Q_c^{-1} + Q_0^{-1} + Q_{MB}^{-1}$. Here $Q_c$ is the coupling quality factor between the KID and the feedline, and $Q_0$ is due to loss mechanisms besides quasiparticles. $y$ is the frequency detuning after taking into account the frequency shift caused by nonlinear kinetic inductance. $y$ depends on a parameter $a$ expressing the strength of the nonlinearity: $y=Q_rx+a/(1+4y^2)$ \cite{swenson}. Here, the total detuning prior to calculating the frequency shift due to nonlinearity is $x=x_0 + \delta x_{MB}$, where $x_0$ is the fractional detuning between the probe tone and a fiducial resonant frequency: $x_0 = (f_{tone}-f_r)/f_r$. $f_r$ would be the resonant frequency of the KID if $a=0$ and $\delta n_{qp} = 0$.

\begin{figure}[htbp]
\centering
\includegraphics[width=\linewidth, keepaspectratio]{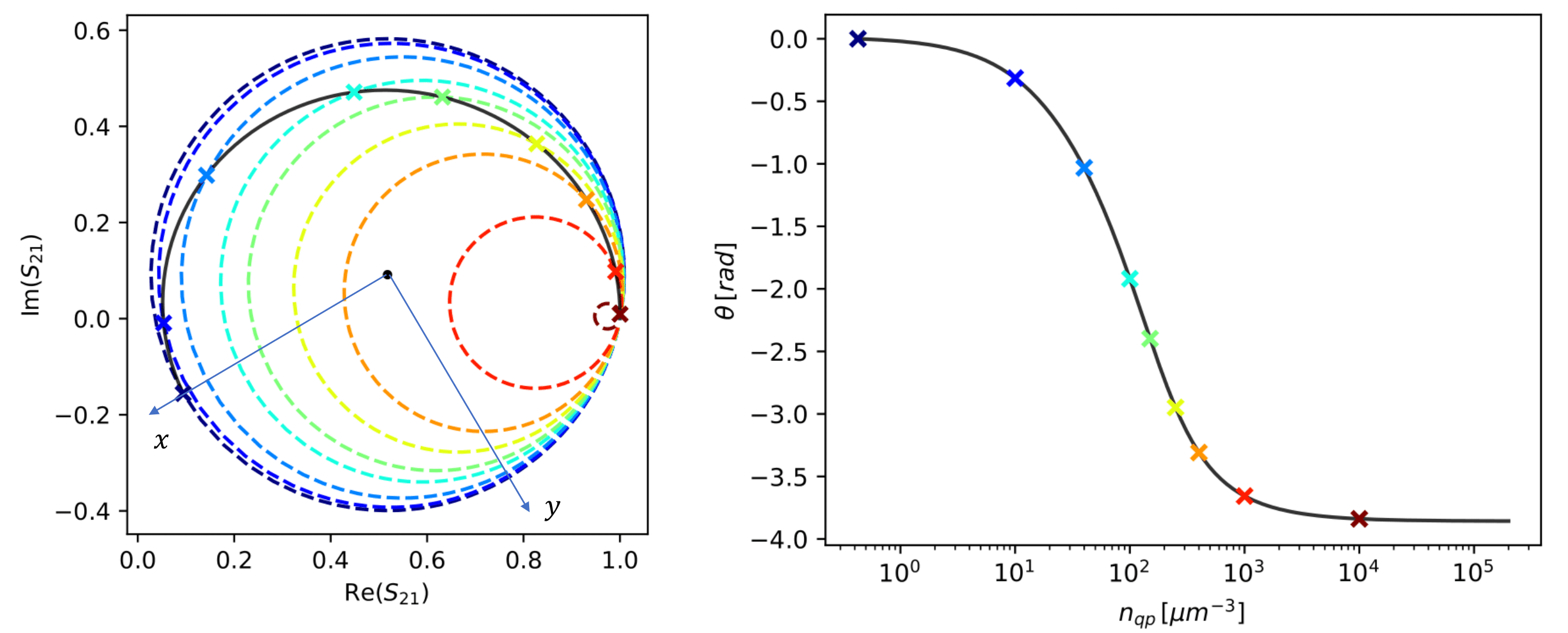}
\caption{{\it Left:} Dependence of the microwave transmission $S_{21}$ on quasiparticle density and microwave frequency. The color circles represent a probe tone sweep taken at different quasiparticle densities, and the black line represents a fixed probe tone as the quasiparticle density is increased. {\it Right:} Plot of the transmission angle vs. the quasiparticle density.}
\label{fig:nqpTheta}
\end{figure}

If we sweep the probe tone's frequency at a fixed density $n_{qp}$, with $\delta n_{qp}=0$, then $y$ will vary through $x_0$, and $S_{21}$ will trace out a circle in the complex plane. We will adopt a coordinate system with its origin at the center of this circle, as shown in the left plot of Fig. \ref{fig:nqpTheta}. Note that the smaller circles represent the same sweep being performed at increasing values of $\delta n_{qp}>0$. Now consider fixing the probe tone frequency and adding quasiparticles to the system ($\delta n_{qp}>0)$. The trajectory of $S_{21}$ follows the black line in the left plot of Fig. \ref{fig:nqpTheta}. If we now calculate the angle $\theta = Re(S_{21})/Im(S_{21})$ with respect to our new coordinate system, then $\theta$ is a monotonically decreasing function of $n_{qp}$. This fact can be used to infer the quasiparticle density from a measurement of $S_{21}$.

\section{Single-Photon Measurements}
\label{sec:photon}

We start by selecting a KID on the $\lambda=25\mu m$ device. The device is cooled to a temperature of 100mK. In a dark configuration, we perform a frequency sweep of the resonance and fit the $S_{21}$ of this sweep to Eq. \ref{eq:S21}, with the fit parameters $(f_r, Q_r, Q_c, \phi, a)$. $Q_{MB}$ is calculated using Eq. (\ref{eq:MB}), with the quasiparticle density at its thermal equilibrium value: $n_{qp} = n_{th} = 2N_0\sqrt{2\pi k_bT\Delta_0}\exp(-\Delta_0/k_bT)$. Based on electromagnetic simulations, we take $\alpha_{kin}=0.8$. We use a density of states $N_0=1.72\times 10^{10}\mu m^{-3}eV^{-1}$. Prior measurements of this detector across a range of temperatures have suggested a best fit value of $T_c = 1.36K$, which yields a gap energy of $\Delta_0 = 1.76k_bT_c = 0.206$meV. $Q_0$ is then calculated as $Q_0^{-1}=Q_r^{-1}-Q_c^{-1}-Q_{MB}^{-1}$. 

We illuminate the device with a near-IR laser $(1550\pm 50 nm)$ set to produce single-photon pulses. We measure a continuous timestream with a sample rate of 200kHz. The photon energy is sufficiently high that the pulses can be easily distinguished from the noise, and 873 pulses were identified in a 200 second timestream. Using an optimal filter \cite{sunil}, we line up the start times of the pulses and average them to obtain a template pulse which is assumed to be a noiseless version of a photon event. The IQ circle fit is used to convert $S_{21}$ to $\theta$ as described in Sec. \ref{sec:background}. With the IQ fit and material parameters listed in the above paragraph, we tabulate a list of $\theta(n_{qp})$ using Eqs. (\ref{eq:MB}) and (\ref{eq:S21}). Since $\theta(n_{qp})$ is monotonic, it can be easily inverted to obtain $n_{qp}$ at each point of the template pulse. Finally, we obtain the perturbation to the quasiparticle number as $\delta N_{qp} = V(n_{qp}-n_{th})$, where the inductor volume is $V=18 \mu m^3$.
\begin{figure}[htbp]
\centering
\includegraphics[width=\linewidth, keepaspectratio]{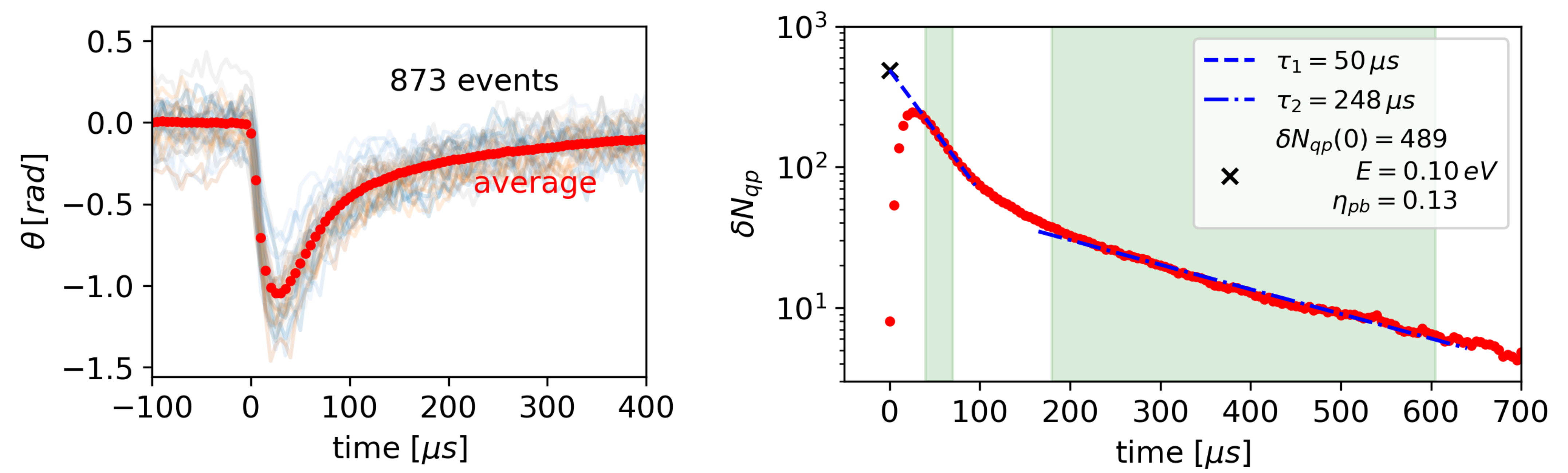}
\caption{{\it Left:} Plot of all photon events with the average overlaid in red. {\it Right:} Plot of the averaged event converted to quasiparticle number, with the fits to the fast and slow decays overplotted. The shaded regions indicate the data over which the fits were performed.}
\label{fig:photonPulse}
\end{figure}

In the pulse profile of $\delta N_{qp}$ (Fig. \ref{fig:photonPulse}, right), we see a resonator ring-up at the start, lasting $\sim 25\mu s$. This matches what we would predict from our IQ circle fit: with fit parameters $f_r = 751.5$ MHz and $Q_r=61900$, we obtain $\tau_{ring}=f_r/(\pi Q_r) = 26\mu s.$ The ring-up is followed by a fast decay and then a slower decay. We fit each decay with an exponential model, giving lifetimes of $50\mu s$ and $248\mu s$ respectively. The second, longer lifetime is taken to be the near-equilibrium quasiparticle lifetime: $\tau_{qp}\sim 250\mu s$.

When Cooper pairs are broken by a near-IR photon, the resulting high-energy quasiparticles relax back to energies close to the gap on the timescale of 0.1-10 ns \cite{downconv}. Thus, by the time the first sample is taken, the quasiparticle number $N_{qp}$ has reached its maximum value and is starting to decay due to recombination. The $\sim 25\mu s$ response time of the resonator prevents us from resolving $N_{qp}$ right at the start of the pulse. To overcome this, we extrapolate the first decay back to the start of the event, resulting in an estimate of $\delta N_{qp}(0) \sim 490$. Calculating the pair-breaking efficiency as $\eta_{pb} = \delta N_{qp}(0)\Delta_0/h\nu$, we obtain $\eta_{pb}=0.13$. Note that our measurement is likely still a lower bound, as the quasiparticle recombination may proceed even faster than the exponential fit during the first 25$\mu s$ of the decay due to the increased quasiparticle density at early times.

\section{Cosmic Ray Dead Time}
\label{sec:deadtime}

In L2 orbit, data glitches caused by cosmic ray impacts can result in significant detector dead time. It is necessary to predict this dead time in order to take necessary mitigation measures. First, we estimate the cosmic ray flux in the lab using the same device as in Sec. \ref{sec:photon}. We measure one KID at a time at a sample rate of 200kHz, and convert the timestream of $S_{21}$ to fractional frequency shifts $df/f$ using a calibration from a frequency sweep of the resonance. Peaks in the timestream which rise above $5\sigma$ are flagged as glitches. Two KIDs were read out for $\sim 3500$ seconds each over the course of a day, yielding glitch rates of 5.51/min and 5.53/min.

Measurements of a large array with a Si substrate of similar thickness to ours have indicated that cosmic rays cause $5\sigma$ glitches in KIDs within an area of $\sim 15$ cm$^2$ \cite{karatsu}. Since the area of the chip on which our device is patterned is 1.37 cm$^2$, we can assume that each KID on our small device sees every cosmic ray. Thus, we calculate the cosmic ray flux as (per-pixel glitch rate)$/$(area) $=$ 4/min/cm$^2$, in broad agreement with other measurements \cite{karatsu}. This flux includes both atmospheric muons ($\sim$1/min/cm$^2$) and gamma rays due to radioactive thorium in earth $(\sim 3-6$/min/cm$^2)$. Both types of events deposit on the order of 100keV of energy into the silicon wafer, producing similar responses in a KID. For conciseness, we will refer to both types of events as `cosmic rays'.

The Planck experiment found an energetic particle flux of 300/min/cm$^2$ in L2 orbit \cite{planck}. Thus, we can estimate our expected per-pixel glitch rates at L2 by scaling our rates measured in the lab up by 75$\times$. Note that the device we have presented thus far is much smaller in area than the arrays that will be deployed at L2. As the array size increases, more cosmic rays will pass through the substrate, causing the per-pixel glitch rate to increase. Thus, a mitigation strategy for cosmic rays is necessary with larger wafers. Based on the work of K. Karatsu et. al. \cite{karatsu}, a low-Tc Ti layer has been added to our larger arrays in order to downconvert non-thermal phonons to sub-gap energies, decreasing the area around each KID which will cause a glitch due to a cosmic ray impact. This layer has been implemented in a PRIMA FIRESS prototype array with 1008 pixels and an area of 10.3 cm$^2$.

\begin{figure}[htbp]
\centering 
\includegraphics[width=\linewidth, keepaspectratio]{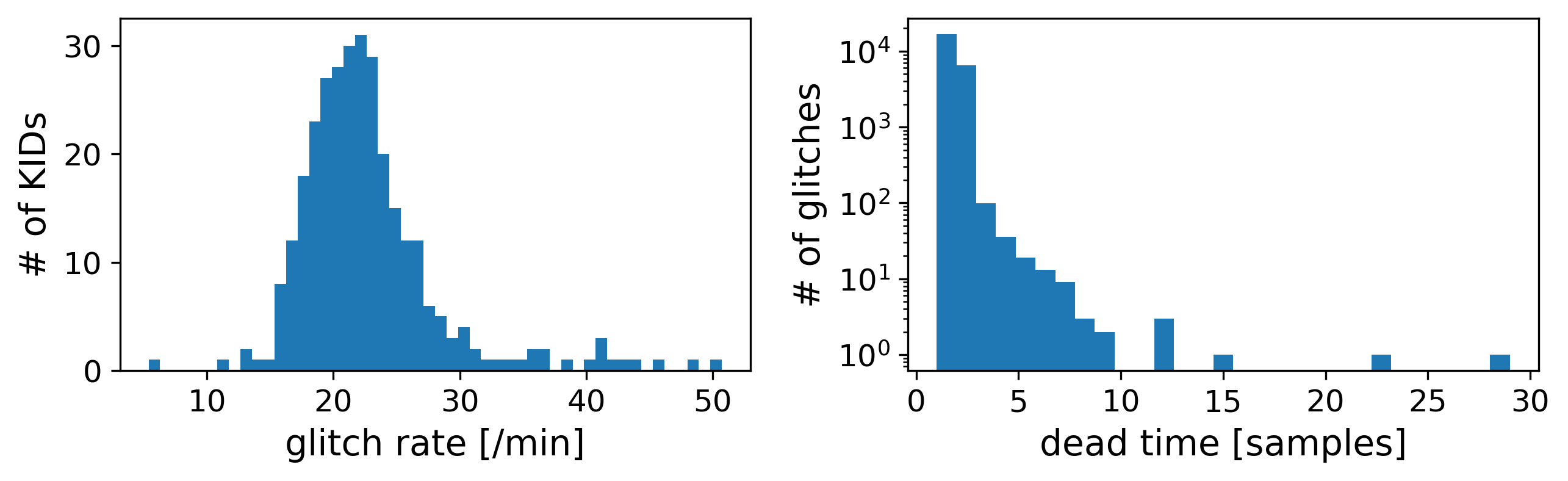}
\caption{{\it Left:} Histogram of per-pixel cosmic ray rates for 310 KIDs. {\it Right:} Histogram of dead times from each individual glitch across all 310 KIDs. The time per sample is 0.00205 sec.}
\label{fig:cr_hists}
\end{figure}

Out of 1008 pixels, 772 yielded and 462 of the remaining KIDs were rejected due to frequency collisions, misplaced tone frequency, or unoptimized tone power, leaving 310 KIDs for analysis. To perform multi-tone measurements, we use a Xilinx Radio Frequency System on a Chip (RFSoc) with firmware \cite{firmware} and software \cite{software} developed for the Prime-Cam instrument \cite{primecam}. The readout was performed at a sample rate of 488 Hz for 200 seconds.  Glitches were identified as peaks with a signal of $5\sigma$ or greater. For each glitch, the dead time was defined as the time for the signal to return to the $5\sigma$ level. The results are presented in Fig. \ref{fig:cr_hists}. The mean per-pixel glitch rate is $R_{pix}=0.38/$sec and the mean dead time per glitch is $t_{dead}=0.0027$ sec. Thus, the projected dead time fraction at L2 would be $R_{pix}\times t_{dead}\times 75 = 8\%$.

The per-pixel glitch rate is higher than expected with the low-Tc layer in place, but analysis of the multiplicity (number of KIDs that experience a glitch simultanously) shows that most glitches are likely not caused by cosmic rays. Each time a glitch was detected in any KID, a glitch in any other KID within the next 3 samples was conservatively assumed to be caused by the same cosmic ray. Through this process, we identified $\sim 12000$ cosmic rays in 200 seconds, most of which have multiplicity $=1$. This cosmic ray rate would suggest an unreasonably high flux of $\sim 350$/min/cm$^2$. We conclude that there is an excess of low-multiplicity glitches which are not caused by cosmic rays.

\begin{figure}[htbp]
\centering
\includegraphics[width=\linewidth, keepaspectratio]{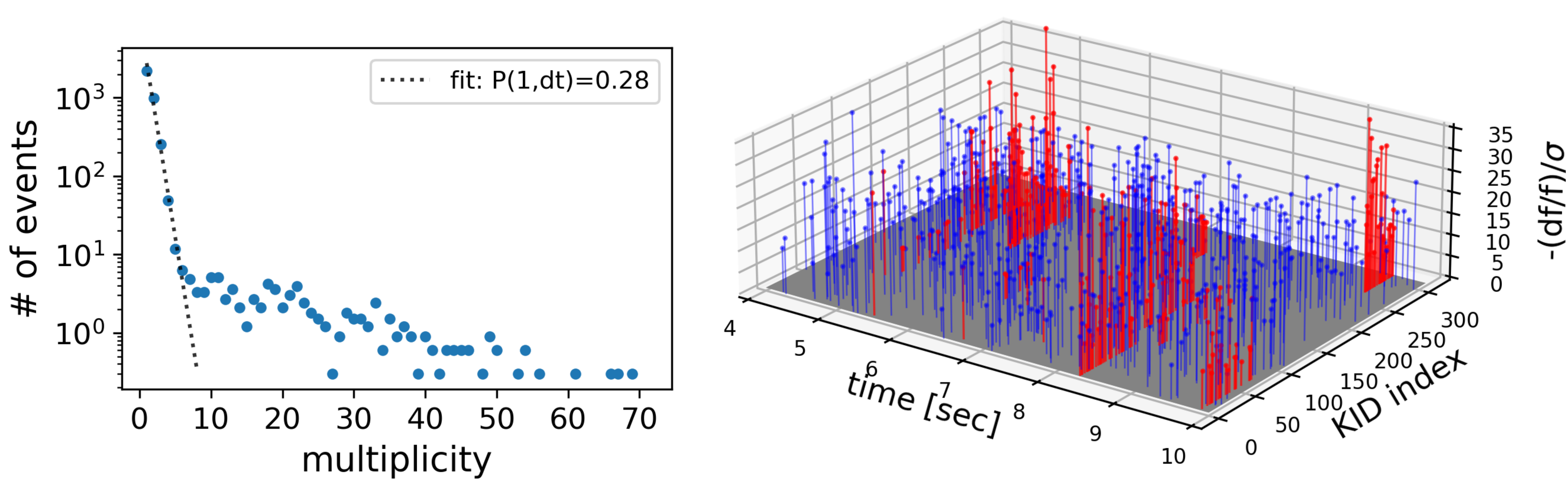}
\caption{{\it Left:} Number of events per multiplicity. Multiplicity denotes the number of KIDs affected at the $5\sigma$ level by an event. {\it Right:} Typical 6 second time trace of all 310 KIDs. Each trace is normalized along the z-axis by $\sigma$. Events with multiplicity $>$ 4 are highlighted in red.}
\label{fig:mult}
\end{figure}

In Fig. \ref{fig:mult}, we see that the event counts at low multiplicity fit well to a power law $N(m)=N(0)p^m$, with $m$ the multiplicity. This suggests that low-multiplicity events are produced by a process individually acting on each pixel with a constant rate, since such a process would contribute a fixed conditional probability of an additional pulse appearing in a different KID within 3 samples after the first pulse is detected. The power-law behavior is observed to stop at a multiplicity of 5, so we assume that cosmic rays become dominant here and remove all events with multiplicity less than 5. By calculating the per-pixel rate after these events were removed, we obtain a value of 0.1/sec. This yields a lower estimate of $2\%$ dead time at L2. After removing the low-multiplicity events, there are still $\sim 340$ cosmic rays detected, giving an estimated flux of $\sim$10/min/cm$^2$. This is greater than our earlier estimate of 4/min/cm$^2$, indicating that there may still be an additional source of events at somewhat higher multiplicity than 5.

\section{Summary and Discussion}

We have presented measurements of the pair-breaking efficiency and quasiparticle lifetime of a KID on a small array designed for observations at $\lambda=25\mu m$. Using single-photon excitations, we measure a lifetime of $\tau_{qp}\sim 250\mu s$ at $T=100$mK and a pair-breaking efficiency of $\eta_{pb}\geq 0.13$. We stress that our estimate of $\eta_{pb}$ is a lower bound, due to the detector's finite response time obscuring the earliest parts of the decay. This estimate could be improved using a detector with a lower coupling quality factor $Q_c$ to decrease the response time. 

We constrain the cosmic ray dead time at L2 to be between $2\%$ and $8\%$ for a FIRESS prototype array designed for observations at $\lambda=210\mu m$. These two estimates are both below the $10\%$ rate adopted in PRIMA's instrument sensitivity model. We concluded that most of the low-multiplicity events included in the $8\%$ upper bound are likely not cosmic rays, as cosmic rays should not primarily appear in only one or two detectors at a time. Future work will be dedicated towards understanding the source of the low-multiplicity events, eliminating them, and performing a more precise measurement of the cosmic ray dead time.

\backmatter

\bmhead{Acknowledgments}

The research was carried out at the Jet Propulsion Laboratory, California Institute of Technology, under a contract with the National Aeronautics and Space Administration (80NM0018F0610). This work was funded by the NASA (Award No. 141108.04.02.01.36)---to Dr. C. M. Bradford. We would like to thank Thomas Stevenson for helpful discussions on the source of gamma rays in the lab.


\end{document}